# A general cipher for individual data anonymization

Nicolas Ruiz[1]
OECD

**Abstract**

Over the years, the literature on individual data anonymization has burgeoned in many directions. Borrowing from several areas of other sciences, the current diversity of concepts, models and tools available contributes to understanding and fostering individual data dissemination in a privacy-preserving way, as well as unleashing new sources of information for the benefits of society at large. However, such diversity doesn't come without some difficulties. Currently, the task of selecting the optimal analytical environment to conduct anonymization is complicated by the multitude of available choices. Based on recent contributions from the literature and inspired by cryptography, this paper proposes the first cipher for data anonymization. The functioning of this cipher shows that, in fact, every anonymization method can be viewed as a general form of rank swapping with unconstrained permutation structures. Beyond all the currently existing methods that it can mimic, this cipher offers a new way to practice data anonymization, notably by performing anonymization in an ex-ante way, instead of being engaged in several ex-post evaluations and iterations to reach the protection and information properties sought after. Moreover, the properties of this cipher point to some previously unknown general insights into the task of data anonymization considered at a general level of functioning. Finally, and to make the cipher operational, this paper proposes the introduction of permutation menus in data anonymization, where recently developed universal measures of disclosure risk and information loss are used ex-ante for the calibration of permutation keys. To justify the relevance of their uses, a theoretical characterization of these measures is also proposed.

### 1. Introduction and contributions of this paper

Data on individual subjects are increasingly gathered and exchanged. By their nature, they provide a rich amount of information that can inform statistical and policy analysis in a meaningful way. However, due to the legal obligations surrounding these data, this wealth of information is often not fully exploited in order to protect the confidentiality of respondents and to avoid privacy threats. In fact, such requirements shape the dissemination policy of individual data at national and international levels. The issue is how to ensure a sufficient level of data protection to meet releasers' concerns in terms of legal and ethical requirements, while still offering users a reasonable level of information. Over the last decade the role of micro data has changed from being the preserve of National Statistical Offices and government departments to being a vital tool for a wide range of analysts trying to understand both social and economic phenomena. This has raised a new range of questions and pressing needs about the privacy/information trade-off and the quest for best practices that can be both useful to users but also respectful of respondents' privacy.

Statistical disclosure control (SDC) research has a rich history of addressing those issues by providing the analytical apparatus through which the privacy/information trade-off can be assessed and implemented. SDC consists in the set of tools that can enhance the level of confidentiality of any data while preserving to a lesser or greater extent its level of information (see [1] for an authoritative

---

[1] Contact : nicolas.ruiz@oecd.org. OECD, 2 rue André Pascal, 75016, Paris, France. Tél.: +33145241433

survey). Over the years, it has burgeoned in many directions. In particular, techniques applicable to micro data, which are the focus of this paper, offer a wide variety of tools to protect the confidentiality of respondents while maximizing the information content of the data released, for the benefits of society at large. Such diversity is undoubtedly useful but has, however, one major drawback: a lack of agreement and clarity on the appropriate choice of tools in a given context, and as a consequence, a lack of a comprehensive view (or at best an incomplete one) across the relative performances of the techniques available. The practical scope of current micro data masking methods is not fully exploited precisely because there is no overarching framework. All methods generally carry their own analytical environment, underlying approach and definitions of privacy and information.

A step toward the resolution of this limitation has been recently proposed ([2], [3]), by establishing that any micro data masking method can be viewed as functionally equivalent to a permutation of the original data, plus eventually a small noise addition. This insight, called the permutation paradigm, unambiguously establishes a common ground upon which any anonymization method can be used. However, this paradigm was not originally considered by its author as a new anonymization method *per se*, but instead as a way to evaluate any method applied to any data set on common grounds. This statement can be reconsidered. As will be discussed in this paper, the fact that it provides a post-anonymization common ground for evaluation makes it also suitable for an ex-ante approach to data anonymization, where in fact anonymization is performed directly from permutation. This is the objective of this paper, which develops a new approach to data anonymization by proposing a general cipher based on permutation keys, and which appears to be equivalent to a general form of rank swapping ([4], [5]). Beyond the existing methods that this cipher can universally reproduce, it also offers a new way to practice data anonymization based on the exploration of different permutation structures. This cipher can be used to perform anonymization in an ex-ante way instead of being engaged in several ex-post evaluations and iterations to reach the protection and information properties sought after. The subsequent study of the cipher's properties additionally reveals some new insights into what is the task of anonymization taken at a general level of functioning. Finally, to make this cipher operational, this paper proposes the introduction of permutation menus in data anonymization, where recently proposed universal measures of disclosure risk and information loss are used ex-ante for the calibration of permutation keys. To justify the relevance of their use, a theoretical characterization of these measures is also proposed.

The contributions of this paper are thus several. Following a discussion of some background concepts in Section 2, we define in Section 3 a general cipher for data anonymization, which puts the permutation paradigm on formal grounds. To the best of the author's knowledge, this is the first time a cipher is proposed in the context of individual data anonymization. Because it is able to replicate the outcomes of any anonymization methods, the subsequent study of its properties reveals some previously unknown general principles driving data anonymization. In particular, a pivotal one will turn out to be that the task of anonymization can be performed independently of the data to be anonymized. We thus argue in Section 4 that this principle paves a new approach to data anonymization. Instead of going through a succession of trial and errors to reach the desired levels of utility and risk before releasing a data set, which is the way most of the current methods are put into practice ([6]), data anonymization can in fact be performed through an ex-ante approach. The use of suitable metrics, that will be characterized theoretically, allows calibrating the keys of the cipher and then anonymizing the data with some pre-defined levels of risk and utility that will be automatically translated onto the anonymized data. The ex-post checking of the properties of the anonymized data set thus becomes irrelevant, which delivers a more efficient approach for conducting anonymization. These results open several new lines of research gathered in Section 5.

2. **Background and related work**

In this section, we summarize the general approach to individual data transaction and the basics of SDC methods, as well as their commonalities and differences with cryptography. We then turn to the description of the recent functional equivalence in anonymization for ex-post evaluation established by the permutation paradigm, upon which this paper is based.

## 2.1 Individual data transaction and protection

A general and standard way of describing a transaction on individual data is to consider two agents: a data releaser, which supplies individual data, e.g. public administrations, enterprises, and some data users, who demand individual data, e.g. researchers, public administrations, enterprises. The former typically gathers, under some suitable forms, a micro data set that is data collected from individuals. The latter will have various needs in terms of information and seeks the data in order to conduct a potentially large variety of tasks. Note that in this simple setting we assume trustworthiness on the supplier side, meaning that the data releaser knows the identities of the respondents who contributed to the data set out of their good will. Moreover, we do not restrict the set of potential tasks to be conducted by the data users, which thus can range from simple data mining tasks such as frequencies counts and computation of the mean and median of a distribution, to more elaborate tasks such as econometric techniques. Alternatively, this is equivalent to considering that the data releaser is not equipped with sufficient technical knowledge to conduct the different tasks that the users have in mind. Thus, data are released without being tailored to very specific needs.

The delivery of the micro data set by the data releaser to the data users, via any potential channel, is what characterizes a transaction on individual data. The users then go away with the data to perform some tasks on them without any further interaction with the releaser. As such, and as previously defined in the literature, the transaction is a standard non-interactive one ([7]). Naturally, other types of transactions are possible: for example, and by assuming that the data releaser has sufficient technical knowledge, data mining tasks could be performed on the data by the releaser upon request of the users, and the former will communicate the outputs of the tasks to the latter. For such an interactive transaction, differential privacy has gained strong momentum in the literature to conceptualize and tackle the issues that could arise in terms of privacy protection. However, some questions remain unresolved, such as the quality of the output that is delivered to the users in terms of information ([8]). Moreover, and because in an interactive transaction a mechanism is sitting between the releaser and the users in order to perform the tasks, it is ultimately outputs that are delivered to the users, not data *per se*. As a result, one has to make some untenable assumptions about the users' needs, by inevitably restraining them or similarly assuming a very expert data releaser that could perform any kind of task. As noted, this is not what we will assume in this paper because such assumptions could only lead to unrealistic, or at most, specific forms of data transactions[2]. Given these limitations, the scope of this paper is thus voluntarily narrowed to the non-interactive exchange of data sets.

Non-interactive data transactions immediately raise the pressing question of privacy, even more so than in other forms of exchange. In modern societies with pervasive data collection, it is a matter of general interest to grant access to individual data, but not to the detriment of privacy, a fundamental right for all individuals. The exchange of individual data in their original form, as collected by the releaser, generally entails a violation of individual privacy given the sensitive information that they can contain. This is why the privacy legislations that prevail in most countries preclude the dissemination of data that are linkable to individuals or allow the recovery of some of their characteristics. So, in order to prevent any disclosure on individuals, data have first to be anonymized through the application of suitable statistical disclosure control (SDC) techniques.

SDC research has a long and rich history in providing data releasers with a set of tools for anonymizing individual data under various settings ([1]). In a nutshell, for non-interactive data exchange, the overall approach of SDC is for a data releaser to modify the original data set in some way that reduces disclosure risk while altering as little as possible the information that it contains. As a result, at a general level SDC techniques can be classified into two main approaches:

---

[2] For example, beyond simple mining tasks it remains currently unknown if the strong privacy guarantees offered by differential privacy can be translated to elaborated econometric works, thus *de facto* excluding users with such needs.

- *Privacy-first*: the method is applied with the primary goal of complying with some pre-requisites on the level of privacy, judged as acceptable and under which data exchange can take place.
- *Utility-first*: the method is applied with the primary goal of complying with some pre-requisites on the level of information judged as valuable enough to make data exchange worthwhile.

The privacy-first approach shares certain features with cryptography. Indeed, the act of protecting privacy through anonymization can be conceived as a form of encryption, where it is the individuals' identities that are encrypted. However, the utility-first approach establishes a first fundamental difference with cryptography, as it would be pointless to release micro data that contain no information at all. So, while the very goal of cryptography is to release a cypher text that discloses nothing whatsoever about the underlying plaintext, the purpose of individual data exchange is to release data (i.e. the cypher text) considered as safe as possible in term of privacy, while purposefully leaking some information (and generally, the more the better). A second fundamental difference lies in the types of agents involved and how the transaction operates.

In cryptography, a sender encrypts a message and the receiver decrypts it with the appropriate key, while in the middle an attacker tries to intercept the message and to decipher it using cryptanalysis techniques. In an individual data exchange, first, there is ideally no decryption phase: the data user takes the released data set as given for his analysis needs. Second, while in cryptography there is a clear distinction between sender, receiver and attacker, in an individual data exchange the receiver can also be an attacker. Indeed, a malevolent user could potentially try to re-identify individuals in a data set and the data releaser has no way of preventing this after the exchange takes place (nor would an ex-ante screening of the users to identify the reliable ones preclude, in principle, that they become attackers). Finally, a third difference is that the re-identification of individuals, which constitutes an attack in data anonymization, carries a different meaning than an attack in cryptography. Indeed, while in the latter case the single objective is generally to retrieve the full plaintext, in the former this not so: the re-identification of at least one individual can already be considered as a successful attack. Thus, the cryptographic viewpoint of an attack in data anonymization is about identifying some individuals but not necessarily all of them, i.e. some of the plaintext but not necessarily all of it. To summarize, while in principle micro data are not meant to be deciphered, the releaser must sufficiently encipher the data so as to prevent any re-identification of individuals, while at the same time ensuring that the data contain a sufficient level of information to be meaningful to most users. Here lies the fundamental trade-off in individual data exchange that is not present in cryptography: encryption, i.e. privacy preservation, *versus* information leakage. The goal of SDC techniques is to manage this trade-off in a meaningful and practical way.

To do so, research in SDC has led to a wide variety of tools, spanning several possible types of data across different fields. This diversity of available techniques is undoubtedly an asset but it comes also with some drawbacks ([9]). The lack of an overarching framework upon which the trade-off between utility and disclosure risk can be assessed is problematic because it leads to an absence of consensus regarding "best practices". In fact, the current state of the literature, while of high quality, offers at best techniques tied to the context upon which they operate. For example, comparing the level of utility and privacy achieved by different methods on different data sets is an awkward task as different metrics and/or different parametrizations are largely heterogeneous, so that no common ground exists for comparison. This is why generally only ad-hoc comparisons can be conducted ([3]). Additionally, each metric embodies distributional dependence and this feature has a significant impact for the performance evaluation of masking methods across data sets ([9]).

Another issue stems from the fact that agents involved in a data exchange all have different views regarding privacy and utility. Typically, different releasers will have different views on what is an acceptable level of privacy, and users will overall tend to put greater weight on data utility and be less concerned about privacy than most releasers. Moreover, inside the population of users, information needs can greatly differ according to what they plan to do with the data. Currently, SDC techniques don't offer sufficient flexibility to account for such heterogeneity in preferences. The above distinction between utility and privacy-first approaches is undoubtedly a way toward the representation of different sensitivities, but it nonetheless remains at a high level of principle and is also barely met in practice. In fact, in a utility (resp. privacy)-first approach, it is advisable to check the value of privacy (resp. utility) achieved by the methods before data dissemination, which thus always lead to the limitation of context-dependence exposed above. Moreover, inside each approach, none of the methods available can straightforwardly cope with different preferences in terms of privacy and information.

To address these issues and the need to generalize the concepts used in SDC, a recent contribution in the literature proposed a general functional equivalence based on permutations to describe any data masking method (see [3] and its subsequent development in [2]).

### 2.2 The permutation paradigm in data anonymization

The permutation paradigm in data anonymization starts from the observation that any anonymized data set can be viewed as a permutation of the original data plus a non-rank perturbative noise addition. Thus, it establishes that all masking methods can be thought of in terms of a single ingredient, i.e. permutation. This result clearly has far reaching conceptual and practical consequences, in the sense that it provides a single and easily understandable reading key, independent of the model parameters, the risk measures or the specific characteristics of the data, to interpret the utility/protection outcome of an anonymization procedure.

To illustrate this equivalence, we introduce a toy example which consists (without loss of generality) of five records and three attributes $X=(X_1, X_2, X_3)$ generated by sampling $N(10,10^2)$, $N(100,40^2)$ and $N(1000,2000^2)$ distributions, respectively. Noise is then added to obtain $Y=(Y_1, Y_2, Y_3)$, the three anonymized version of the attributes, from $N(0,5^2)$, $N(0,20^2)$ and $N(0,1000^2)$ distributions, respectively. One can see that the masking procedure generates a permutation of the records of the original data (Table 1).

**Table 1. An illustration of the permutation paradigm**

| Original dataset X | | | Masked dataset Y | | |
|---|---|---|---|---|---|
| $X_1$ | $X_2$ | $X_3$ | $Y_1$ | $Y_2$ | $Y_3$ |
| 13 | 135 | 3707 | 8 | 160 | 3248 |
| 20 | 52 | 826 | 20 | 57 | 822 |
| 2 | 123 | -1317 | -1 | 122 | 248 |
| 15 | 165 | 2419 | 18 | 135 | 597 |
| 29 | 160 | -1008 | 29 | 164 | -1927 |
| Rank of the original attribute | | | Rank of the masked attribute | | |
| $X_{1R}$ | $X_{2R}$ | $X_{3R}$ | $Y_{1R}$ | $Y_{2R}$ | $Y_{3R}$ |
| 4 | 3 | 1 | 4 | 2 | 1 |
| 2 | 5 | 3 | 2 | 5 | 2 |
| 5 | 4 | 5 | 5 | 4 | 4 |
| 3 | 1 | 2 | 3 | 3 | 3 |
| 1 | 2 | 4 | 1 | 1 | 5 |

Now, as long as the attributes' values of a data set can be ranked, which is obvious in the case of numerical and categorical ordinal attributes, but also feasible in the case of nominal ones ([10]), it is always possible to derive a data set Z that contains the attributes $X_1$, $X_2$ and $X_3$, but ordered according to the ranks of $Y_1$, $Y_2$ and $Y_3$, respectively, i.e. in Table 1 re-ordering $(X_1, X_2, X_3)$ according to $(Y_{1R}, Y_{2R}, Y_{3R})$. This can be done following the post-masking reverse procedure outlined in ([3]). Finally, the masked data Y can be fully reconstituted by adding small noises $(E_1, E_2, E_3)$ (small in the sense that they cannot re-rank Z while they can still be large in absolute values) to each observation in each attribute (Table 2).

**Table 2. Equivalence in anonymisation: postmasking reverse mapping plus noise addition**

| Original dataset X | | | Reverse mapped dataset Z | | |
|---|---|---|---|---|---|
| $X_1$ | $X_2$ | $X_3$ | $Z_1$ | $Z_2$ | $Z_3$ |
| 13 | 135 | 3707 | 13 | 160 | 3707 |
| 20 | 52 | 826 | 20 | 52 | 2419 |
| 2 | 123 | -1317 | 2 | 123 | -1008 |
| 15 | 165 | 2419 | 15 | 135 | 826 |
| 29 | 160 | -1008 | 29 | 165 | -1317 |

| Noise E | | | Masked dataset Y(=Z+E) | | |
|---|---|---|---|---|---|
| $E_1$ | $E_2$ | $E_3$ | $Y_1$ | $Y_2$ | $Y_3$ |
| -5 | 0 | -459 | 8 | 160 | 3248 |
| 0 | 5 | -1597 | 20 | 57 | 822 |
| -3 | 0 | 1256 | -1 | 122 | 248 |
| 2 | 0 | -229 | 18 | 135 | 597 |
| 0 | -1 | -610 | 29 | 164 | -1927 |

By construction, Z has the same marginal distributions as X, which is an appealing property. Moreover, under a maximum-knowledge intruder model of disclosure risk evaluation, the small noise addition turns out to be irrelevant ([2] and Section 3): re-identification *via* record linkage can only come from permutation, as by construction noise addition cannot alter ranks. Reverse mapping thus establishes permutation as the overarching principle of data anonymization, allowing the functioning of any method to be viewed as the outcome of a permutation of the original data, independently of how the method operates. This result has been explicitly proposed by its authors for the ex-post evaluation of anonymization, but not as a new technique for conducing anonymization. As we will see, it can in fact be viewed and operationalized as a new, general framework for anonymization.

One may be surprised by this result and ask why all in all, the fundamental principle of data anonymization appears to be as simple as permutation. After all, in cryptography, permutation cyphers and their cryptanalyses have been known for centuries. They are easy to detect because they don't affect individual symbols' frequencies (the equivalent of this in the permutation paradigm being the preservation of marginal distributions). In fact, and as will be discussed in what follows, it turns out that the obvious weakness of permutation cypher in standard cryptography shows up as strength in data anonymization, in the sense that the degree of permutation performed allows to control for the amount of information that is leaked. Moreover, because the permutation paradigm proposes one single universal language for data anonymization, it allows introducing some measures of disclosure risk and information loss that can be used in any context, and flexible enough to capture the variety of views that can occur in a data exchange ([11] and Section 4 below). While these measures were also originally proposed for the ex-post evaluation of the outcomes of any anonymization techniques on any data, they can in fact be used equally validly ex-ante to conduct anonymization.

### 3. Definition and properties of a cipher for data anonymization

### 3.1 Data anonymization as a cipher

We first recall a proposition that has been made in a recent contribution and that naturally results from the permutation paradigm ([11]):

**Proposition 1**: *For a data set[3] $X_{(n,p)}$ with n records and p attributes ($X_1,..,X_p$), its anonymized version $Y_{(n,p)}$ can always be written, regardless of the anonymization methods used, as:*

$$Y_{(n,p)} = \left(P_1 X_1, \ldots, P_p X_p\right)_{(n,p)} + E_{(n,p)}$$

*where $P_1,..,P_p$ is a set of p permutation matrices and $E_{(n,p)}$ is a matrix of small noises.*

*Proposition 1* highlights the fact that because permutation appears to be the overarching principle ruling data anonymization, the functioning of any method can be expressed as a set of permutation matrices, plus eventually a matrix of small noises. Despite the large heterogeneity in the methods currently available, e.g. rank-based, noise-based, cluster-based, they can in fact essentially all be viewed as the application of permutation matrices to the original data set. This proposition forms the basis upon which a cipher for data anonymization can be built. However, it remains limited in the sense that the permutation keys are not isolated. Indeed, except in the particular case where all the pairwise correlation across the p attributes are equal to one, the set of $P_1,..,P_p$ matrices will not measure the amount of permutation. To do so, each attribute needs first to be sorted in increasing order, which can be viewed as preliminary permutations, then the levels of permutations aimed at anonymizing the data set are introduced, and finally the sorting is undone through the inverse permutation matrix of the first step. This leads to the following proposition:

**Proposition 2**: *For a data set $X_{(n,p)}$ with n records and p attributes ($X_1,..,X_p$), its anonymized version $Y_{(n,p)}$ can always be written, regardless of the anonymization methods used, as:*

$$Y_{(n,p)} = \left(A_1^T D_1 A_1 X_1, \ldots, A_p^T D_p A_p X_p\right)_{(n,p)} + E_{(n,p)}$$

*where $A_1,..,A_p$ is a set of p permutation matrices that sort the attributes in increasing order, $A_1^T,..., A_p^T$ a set of p permutation matrices that put back the attribute in the original order, $D_1,..,D_p$ is a set of permutation matrices for anonymizing the data and $E_{(n,p)}$ is a matrix of small noises.*

*Proposition 2* describes the fundamental functioning of any anonymization method with the permutation keys made explicit. Proceeding attribute by attribute, each is first permuted to appear in increasing order, then the key is injected, and finally it is re-ordered back to its original form by applying the inverse of the first step (which in the case of a permutation matrix is simply its transpose). A small noise is also eventually added. Clearly, we have that $P_j = A_j^T D_j A_j \ \forall j = 1, \ldots, p$ with $D_1,..,D_p$ subsuming the properties of any anonymization method by capturing the amount of permutation performed. For example, considering the following permutation matrix $D_j$ applied to a given attribute j:

$$\begin{pmatrix} 0 & 0 & 0 & 0 & 1 \\ 0 & 1 & 0 & 0 & 0 \\ 0 & 0 & 1 & 0 & 0 \\ 1 & 0 & 0 & 0 & 0 \\ 0 & 0 & 0 & 1 & 0 \end{pmatrix}$$

and counting line by line how this matrix departs from the identity matrix:

$$\begin{pmatrix} 1 & 0 & 0 & 0 & 0 \\ 0 & 1 & 0 & 0 & 0 \\ 0 & 0 & 1 & 0 & 0 \\ 0 & 0 & 0 & 1 & 0 \\ 0 & 0 & 0 & 0 & 1 \end{pmatrix}$$

i.e. how the 1's have been shifted by assigning a negative (resp. positive) sign for a right shifting (resp. left shifting), one can conclude that the first record has been moved 4 ranks down, the fourth 3 ranks up and the fifth 1 rank up, while the second and third records have been left in their original positions. These simple computations are a way of describing the functioning of any anonymization method, but in the language of permutation.

---

[3] In the remainder of this paper, the subscript in parenthesis describes the number of rows and columns for a matrix. Here for example $X_{(n,p)}$ is a matrix with n rows and p columns.

*Proposition 2* thus considers data anonymization at a general level of operation and, following the permutation paradigm, contains all currently existing methods. Interestingly, its nature is similar to the functioning of rank swapping, where data are first sorted in increasing order, permuted within a limited range and then re-ranked according to their original values ([1], [5]). For example, consider the following permutation matrix for one attribute and 6 records:

$$\begin{pmatrix} 0 & 1 & 0 & 0 & 0 & 0 \\ 1 & 0 & 0 & 0 & 0 & 0 \\ 0 & 0 & 0 & 1 & 0 & 0 \\ 0 & 0 & 1 & 0 & 0 & 0 \\ 0 & 0 & 0 & 0 & 0 & 1 \\ 0 & 0 & 0 & 0 & 1 & 0 \end{pmatrix}$$

This matrix, when applied using *Proposition 2*, is a permutation key for rank swapping with a swapping distance equal to one. Thus, data swapping has a functioning that can in fact describe any anonymization method, while it is the swapping distance selected that constrains the structure of the permutation keys. Other methods, such as noise- or cluster-based, will lead to different permutation structure, but ultimately they all boil down to a form of general rank swapping. However, working directly with permutation keys allows uncovering some permutation patterns that may not be mirrored by currently known techniques, which can potentially extend the set of anonymization tools available. This feature will be discussed in more depth in Section 4.

Now that a general key structure has been made explicit, we can define a cipher for data anonymization:

**Proposition 3**: *The three-tuple* $\Gamma = (P, K, E)$ *with the following conditions satisfied:*
- *P is a finite set of possible original and anonymized data sets of $n \geq 2$ records and $p \geq 1$ attributes*
- *K is the keyspace, a finite set of possible key groups k, each containing p permutation-based keys*
- *For each key groups $k \in K$ there exists a group of p permutation-based encryption rules $\varepsilon_k \in E$, where each group $\varepsilon_k: P \to P$ is a function such that $\varepsilon_k(x) = y$ for $\forall x, y \in P$*

*is a cipher for data anonymization.*

This proposition derives from *Proposition 2* and establishes the whole task of data anonymization as a cipher composed of three entities. The first one is the set of possible data sets *P* (i.e. the set of plaintexts in cryptography) of n records and p attributes, e.g. *($X_1$,...,$X_p$)*, which also defines the set of possible anonymized data sets (i.e. the set of cyphertexts). The cipher is thus endomorphic ([12]). It is indeed valid to define a cipher for data anonymization in the particular endomorphic case because, as outlined above, the essential principle of data anonymization is permutation. One can also add some small noises, which are in principle required to recompose *exactly* the outcome of some methods (for example noise-based ones). But the small noises won't change any ranks and thus won't provide any additional protection against disclosure risk. Instead, they will alter the data in a small but unnecessary way that could be detrimental to information. For example, adding small noises won't exactly preserve the marginal distributions of a data set, though such preservation remains a desirable feature of any anonymization tools. Stated otherwise, in data anonymization, it is desirable and somewhat intuitive to expect that any information loss must have as a counterpart improved protection. This is not the case for these non-rank perturbative small noises, as only permutations matter. Consequently, as they don't provide any additional protection but lead instead to superfluous information loss, small noises can be disregarded from the definition of the cipher. Thus, as for permutation ciphers in cryptography, the sets of plaintexts and cyphertexts are the same (while adding small noises would have made the two sets generally different).

The second entity of the cipher is the keyspace *K*. Here, it is important to note that a key is not defined as a single element, which is generally the case in cryptography, but following *Proposition 2*, as a group of p keys, i.e. *($D_1$,...,$D_p$)*, with p being the number of attributes in the data set to be anonymized. Otherwise put, each attribute is equipped with its own key, i.e. a permutation matrix, but this is the group of these p keys that forms the key used for anonymizing the whole data set. As will be made clear in Section 4, the relative properties of the elements within the key group can be used to assess information loss, a feature that differentiates data anonymization from standard cryptography.

Finally, the third element is the set of encryption rules, whereas for the keys an encryption rule is a collection of p specific rules for each attributes. From *Proposition 2*, those rules are given by e.g. $(A_1^T D_1 A_1, \ldots, A_p^T D_p A_p)$, and thus are all based on the products of permutation matrices. However, one crucial departure from standard cryptography is that no decryption rules are postulated and nor are they necessary. As noted in Section 2, individual data exchange doesn't require decryption per se. Once data have been anonymized with the desired levels of disclosure risk and information loss, they are meant to be released and used anonymized. The fact that decryption is not necessary considerably reduces the potential practical difficulties in implementing the cipher. For example, the problem of key exchange as in symmetric-key cryptography doesn't exist here. Moreover, in principle, one doesn't need to select only injective encryption functions to accomplish decryption in an unambiguous manner, albeit in practice it can be noted that because data anonymization relies on permutation the encryption functions will necessarily be injective ([12]). In any case, in the context of data anonymization, this concept appears to be irrelevant.

### 3.2 Some general principles in data anonymization

Having defined a cipher that streamlines the permutation paradigm in data anonymization and that can universally mimic any masking method, we can now characterize some of its properties that will *de facto* pervade to the task of data anonymization in general. We start by a first property that establishes data independence in anonymization:

***Property 1***: *Because it can be defined as a cipher, individual data anonymization can always be performed independently of the data to be anonymized. In particular, the distinction between a utility and privacy-first approach is fundamentally unnecessary.*

The first property is a simple but nonetheless pivotal consequence that stems from the possibility of formulating the task of data anonymization as a cipher. It means that the keys, and thus protection, can be handled and calibrated independently of the data. This may go against the intuition of some statistical disclosure control practitioner, as most of the current existing techniques and their performances are linked to the data upon which they are applied. For example, for multiplicative noise injection with a given parametrization, changes in the distributional characteristics of the data may have a large impact on the level of protection ([13]). More generally, the parameter values of a given method may be a poor indicator of the protection level achieved, as it is the conjunction of these parameters and the distributional characteristics of the data that will ultimately deliver the protection level. As mentioned in the preceding section, this explains why a sequence of trial and error is generally necessary in data masking. Even in a privacy-first approach, ex-post disclosure risk analysis is advised to check if a sufficient level of protection has been effectively achieved. The permutation paradigm, and in this paper its formulation as a cipher, solves this issue, as the permutation keys can be calibrated ex-ante with a given level of protection and thus of information that the encryption will automatically apply to, but independently of, the data. In particular, it turns out that both privacy and utility can be targeted simultaneously and one doesn't have to choose an approach ex-ante and check the other one (or even the two) ex-post.

Originally, the permutation paradigm was proposed to put the comparisons of different methods (and their different parametrizations) across different data sets on a common ground ([3]). Thus, its main goal was the simplification of post-anonymization comparisons. But in fact nothing precludes, conceptually and practically, thinking about data anonymization only in terms of permutations. In return, that means that permutation levels, and thus permutation keys, can be calibrated ex-ante to carry out anonymization instead of being retrieved ex-post to assess the effect of an anonymization method. Thus, whatever the large heterogeneities in the analytical apparatus of masking methods available, they all appear to have an underlying, common permutation-based structure that is independent of the data upon which they are applied.

***Property 2****: Information loss in data anonymization can only come from the alteration of the dependency among attributes, as the cipher $\Gamma$ requires a permutation key per attribute.*

This property narrows the notion of information loss in data anonymization. As mentioned above, given the fact that the overarching principle of data anonymization is permutation, marginal distributions are necessarily always preserved as small noise additions in the reverse mapping procedure are unnecessary. Although they can still be considered, small noise additions are not a fundamental step for recreating the protection outcomes delivered by a method. As a result, the preservation of marginal distributions (non-disclosive in nature), a feature that could appear at first glance as a stringent requirement, is in fact implicitly fulfilled by any anonymization method. This property may also address some recurrent users' concerns about the way data have been modified during the anonymization process, where the addition of noise is sometimes viewed as non-acceptable by some users ([9]). But in fact, any method can ultimately preserve marginal distributions and thus can be analyzed on the anonymized data set in the same way as on the original data. In the cipher $\Gamma$ this fact is made clear by each attribute being equipped with its own permutation key, leaving the attributes' distribution, taken in isolation, unchanged. Information loss can thus only occur from a change in the dependency among attributes, i.e. how attributes will be permuted relative to each other.

***Property 3****: The compounding of two or more anonymization methods is always an inefficient procedure as the cipher $\Gamma$ is idempotent.*

Relying on permutation $\Gamma$ is idempotent, i.e. $\Gamma \times \Gamma = \Gamma$. To see this, assume two unspecified anonymization methods applied sequentially on a given data set. Clearly, each of them has an underlying permutation structure, i.e. they can be expressed respectively as $\Gamma_1 = (P, K_1, E_1)$ and $\Gamma_2 = (P, K_2, E_2)$. The product cipher of $\Gamma_1$ and $\Gamma_2$, denoted $\Gamma_1 \times \Gamma_2$, is defined to be the cipher $(P, K_1 \times K_2, E)$ ([14]). But, the product of two permutation matrices is always a permutation matrix ([15]). Therefore, there is no point in encrypting the data set first with the key $K_1$ and then with $K_2$, as it could have been done directly using a permutation key equal to the product of $K_1$ and $K_2$. In terms of anonymization, that means that compounding two methods necessitates two steps but can't provide more protection than directly using a single step. Instead of targeting a protection level that is known to be reachable by the successive application of two methods (say for example additive noise addition then micro-aggregation), one can calibrate a group of permutation keys to reach this level directly. Consequently, the successive application of different methods is inefficient and anonymization can never reach different outcomes beyond the ones authorized within the set of all permutation keys.

***Property 4****: The cipher $\Gamma$ is pure. Thereby, an adversary attacking an anonymized data set will always face the same kind of cryptanalytic problem, whatever the method used for anonymization.*

Attacks on a data set to re-identify individuals are generally and realistically conceptualized through record linkage, which can be used in the context of any anonymization method and disclosure scenario ([16]). Many different record linkage attacks have been suggested in the literature (see for example ([17]) for an in-depth comparison between distance-based and probability-based procedures), but *Property 4* reduces the type of attacks that can take place on individual data to the same cryptanalytic problem. Because the cipher $\Gamma$ is both endomorphic and idempotent, it is pure. But in a pure cipher, all keys are essentially the same ([14], [18]): whatever key is selected for encryption, an attacker will in fact calculate the same ex-post probabilities of the plaintext. In data anonymization, this translates into the fact that different masking methods ultimately deliver the same kind of challenge for an attacker. Consider for example two arbitrary noise-based and rank-based methods, say additive noise addition and rank swapping. Because additive noise aims at altering the magnitude of the data, one could intuitively think that a distance-based record linkage attack would turn out to be more efficient than a rank-based attack, while the reverse would be true for data swapping. Yet, this is not the case. Because the functioning of any method can always be fundamentally described by an alteration of ranks through a pure cipher, it is ultimately rank-based record linkage attacks that are relevant for both, and in fact, for any anonymization methods.

Indeed, from a heterogeneous selection of methods it has been recently and experimentally characterized in the literature that rank-based record linkage attacks appear to seemingly and consistently outperform distance-based attacks ([19]). While no firm explanation was proposed as to why this is the case, we believe that *Property 4* suggests a response. However, it must be noted that this proposition doesn't convey any additional elements about how to define an adversary, notably which kind of background knowledge he must be empowered with to lead to a reasonable and realistic attack scenario, which is a long-standing issue in the literature ([2]). What *Proposition 4* claims is just that whatever the background knowledge assumed, the cryptanalysis task is always the same and must be based on ranks.

### 3.3 Remarks on the maximum-knowledge attacker model and the validity of the Kerckhoff's principle in data anonymization

The issue of an attacker's background knowledge has been recently pushed further in the literature through the proposal of a maximum-knowledge attacker ([2]), which defines an attacker who knows both the original data set and its entire corresponding anonymized version. This is a rather extreme configuration, unlikely to be mirrored by concrete situations, but it remains however conceptually very insightful, as anonymization that can pass the test of such a situation will in fact be able to pass any test. Note also that this concept provides an additional justification to the irrelevance of small noise additions in data anonymization, as a maximum-knowledge attacker can eliminate the small noise matrix of *Proposition 2* (being able to perform reverse mapping himself), which leaves him to uncover the permutation keys only ([2]).

The concept of a maximum-knowledge attacker is the equivalent of a known-plaintext attack in cryptography. Other types of attack exist but carry less meaning in an individual data exchange. A cyphertext-only attack, where only the anonymized data set is available, is the opposite of a known-plaintext attack, and while the latter may be seen as too stringent, the former is too naïve ([2]). As for chosen plaintext and cyphertext-attacks, they are relevant only in the case in which the attacker can interact with the cipher. Note that a maximum-knowledge attacker, observing both the original data set and its anonymized version, has nothing to gain in terms of information. One can view his attempt as purely slanderous, trying to discredit the data releaser by revealing his permutation keys.

Now, given the assumption that such a powerful person might exist, this leads to one question: given his power, is the task faced by a maximum-knowledge attacker that difficult? In fact, the reply relies on an entity that has not been made explicit in the formulation of the cipher Γ: the record tracking numbers. Generally, data releasers can follow which anonymized record derives from which original record through a number that doesn't carry any information of any sort and is unaffected by encryption. Moreover, when the data are released, all numbers can be modified or deleted. But these numbers, known for practical purposes by the data releaser but not by the maximum-knowledge attacker, act in fact as a mask for the permutation keys. To make this clear, table 3 illustrates the attacker's perspective, using the toy example of section 2.

**Table 3. Point of view of a maximum-knowledge intruder**

| | Original dataset X | | | | Masked dataset Y | | |
|---|---|---|---|---|---|---|---|
| ID | $X_1$ | $X_2$ | $X_3$ | ID | $Y_1$ | $Y_2$ | $Y_3$ |
| 1 | 13 | 135 | 3707 | ■ | 8 | 160 | 3248 |
| 2 | 20 | 52 | 826 | ■ | 20 | 57 | 822 |
| 3 | 2 | 123 | -1317 | ■ | -1 | 122 | 248 |
| 4 | 15 | 165 | 2419 | ■ | 18 | 135 | 597 |
| 5 | 29 | 160 | -1008 | ■ | 29 | 164 | -1927 |

As previously mentioned, it is clear that the attacker can reverse-map the data and eliminate the small noise addition. In this example, he has now to retrieve the permutation key (made of three permutation matrices). In fact, he is already observing some permutation matrices, but those are masked by his ignorance of the tracking numbers, which marks the limit of his knowledge. More explicitly, for each attribute he is observing the product $BA_j^T D_j A_j$: because he has no clue as to who is who between the original and the anonymized data, this is equivalent to assuming that, compared to the data releaser who obviously knows each and every term in the product $BA_j^T D_j A_j$, the attacker is facing an additional, unknown layer of permutation expressed by $B$. Therefore, he is only observing the resulting permutations patterns from the product but not its decomposition. More precisely, despite his knowledge of $A_j$ and its transpose, the matrix $D_j$ that he is trying to recompose is masked by $B$. As $B$ is also a permutation matrix, the attacker is observing an unknown permutation of the encryption keys. As a result, even with his postulated power, due to $B$ the attacker cannot avoid undertaking record linkage because it exists $n!$ possible permutation keys by attributes, and only one will be the correct key.

The fact that the knowledge of the permutation keys will be necessarily hidden when the cipher $\Gamma$ is used makes the Kerckhoff's principle fully relevant in data anonymization ([2]). This principle states that the encryption method must be made available to the public while only the key must be kept secret. In data anonymization, the relevant key ultimately happens to be permutation, no matter how anonymization is practiced. Thus, it can be made public that the cipher $\Gamma$ has been used to protect the data, with the permutation keys remaining secret. Such claim won't weaken the privacy guarantee offered by a data releaser but will contribute to greater clarity in individual data exchange, even in an environment made of maximum-knowledge intruders.

To summarize this section, we formulated data anonymization as an all-purpose cipher that is able to replicate the core functioning of any anonymization method. The formulation in terms of a cipher allows deriving some properties which, while standard in cryptography, when applied in the context of data anonymization, deliver some general guiding principles that, to the best of the author's knowledge, have not been identified so far in the literature. Surely, additional principles could be derived. In particular, one could note that the cipher $\Gamma$ is, theoretically speaking, a one-time pad ([14]). A direct consequence of this is that in principle, perfect secrecy could be achieved in data anonymization ([12]). However, this possibility is a theoretical curiosum which has no empirical validity for at least two reasons which we believe illustrates well the fundamental differences between cryptography and data anonymization. The first is that, as mentioned, the notion of decipherment for individual data is not the same as in cryptography. While in the latter it took place when all the plaintext had been uncovered, in the former it is the amount of correct matches in a record-linkage attack that matters, i.e. which pieces of plaintext have been uncovered, and it doesn't have to be all of them. So even in a one-time pad some correct matches could still be claimed. Thus, the notion of perfect secrecy has no real meaning in data anonymization, except if one makes the requirement that *all* records must be re-identified to qualify a data set as not secure. This is rather unrealistic.

The second reason is that, for $\Gamma$ to be strictly qualified as a one-time pad then the key selection should be truly random. While in cryptography this is fully acceptable, in data anonymization it is not. In addition to providing some privacy guarantees to individuals in the original data, the anonymized data should also meet data users' needs by providing some information. As a result, some structures and constraints must be applied to the permutation keys for the released data to be meaningful. The fact that in data anonymization the keys selection must be guided with both protection and information in mind precludes randomly generating them. In fact, this raises the question of what should be the guidelines to calibrate the keys of the cipher in order to make it concretely usable. This will be discussed in the following section.

### 4. Calibration of the cipher's keys

#### 4.1 Power means in data anonymization

We start by briefly summarizing a recent proposal that postulates power means as an overarching aggregative structure for the computation of universal measures of disclosure risk and information loss ([11]). First of all, it can be observed that for a given attribute j to be encrypted with the cipher and its associated key $D_j$, permutation distances can be retrieved by the computation of a vector of absolute rank displacement $r_j$, i.e. a vector measuring for each record the amount of rank shifting that a key contained[4]. Following what has been developed above, in order to build $r_j$ one can count line by line from the keys how many times the 1s have been moved by using the identity matrix as a benchmark. The evaluation of $r_j$ will lead to an assessment of disclosure risk based on absolute permutation distances.

The evaluation of information loss follows the same approach but based on relative permutation distances. To see this, *Proposition 2* characterized the only possible source of information loss in data anonymization. While marginal distributions can always be preserved, it is the alteration of the dependency among attributes that triggers information loss. In the cipher Γ, this can be captured by the relative properties of the permutation keys applied on two attributes j and j'. Consider for example that the same key is applied to both. Clearly, this is equivalent to permuting a block of attributes and the joint distribution of the block will be exactly preserved. Conversely, once the keys become different, one is necessarily modifying the relative records' positions. This can be captured by a vector $\Delta(r_k)$ of differences between the vectors $r_j$ and $r_{j'}$, which is a vector of dissimilarity between the permutation keys that have been applied to the couple of attributes k=(j, j') (with j≠ j'). When each of the components of $\Delta(r_k)$ are equal to zero (or more precisely to an infinitesimally small value), this depicts the case of a permutation by block of two attributes. But when $\Delta(r_k)$ has some non-zero elements, information has been modified. Note that this approach in terms of information is very general in its scope and relies on a very minimalist assumption: attributes taken pairwise are related by *any* monotic function.

Therefore, disclosure risk and information loss can be evaluated through the use of two related quantities, the distributions of absolute and relative rank displacements. This leads to the following proposition, which establishes power means as a general class of measures for disclosure risk and information loss (see ([11]) for the original and complete proposal):

**Proposition 4**: *Denote by $p=abs(p_1,…,p_n)$ a distribution of permutation distances, being relative or absolute. The following aggregative structure:*

$$J(p,\alpha) = \begin{cases} \left(\frac{1}{n}\sum_{i=1}^{n} p_i^{\alpha}\right)^{\frac{1}{\alpha}} & for\ \alpha \neq 0 \\ \prod_{i=1}^{n} p_i^{\frac{1}{n}} & for\ \alpha = 0 \end{cases}$$

*forms a class of disclosure risk measures when p is a distribution of absolute permutation distances and $\alpha \leq 1$, and a class of information loss measures when p is a distribution of relative permutation distances and $\alpha \geq 1$.*

This proposal makes use of the fact that data anonymization methods all boil down to applying permutation patterns, which greatly simplifies evaluation. When using current methods, protection against disclosure risk and information loss occur at the intersection of two features: the appropriateness and the parametrization of the method selected, and the distributional properties of the data to be anonymized. For instance, if the data are skewed with a heavy tail on the right, then perturbation through additive noise won't provide much protection, generally no matter the amount of noise added ([13]). Indeed, to protect the data sufficiently in that case, a more suitable method, like multiplicative noise, should be selected, and then its correct parametrization to reach the desired levels of protection and information will depend on how the data are effectively skewed. But, when data anonymization is viewed as permutation, then only the alteration of ranks matters. This is why the above proposal depicts a class of universal measures of disclosure risk and information loss, by making abstraction of the interplay between the distributional features of the data and the analytics of the methods. Indeed, the required inputs are only absolute and relative permutation distances.

---

[4] To avoid some unnecessary technical difficulties, zero values in $r_j$ will be assigned, without loss of generality, an infinitesimally small value ε>0.

Using power means also allows capturing the variety of preferences that can occur in a data exchange, through the concepts of aversion to disclosure risk and information loss, something that is new in the literature. The arithmetic mean becomes a special case ($\alpha = 1$) of $J(p, \alpha)$, which forms a natural dividing line by computing the average levels of absolute or relative permutation distances. From this benchmark, and using absolute permutation distances, the more $\alpha$ decreases, the more weight is given to the smallest permutation distances. In fact, the more $\alpha$ approaches -∞, the more $J(p, \alpha)$ converges towards the smallest permutation distance in $r_j$. As a result, for a given $r_j$ and $\alpha' < \alpha$, we have $J(p, \alpha') \leq J(p, \alpha)$ : the lower is $\alpha$, the stronger is the aversion to disclosure risk. Conversely, through the same reasoning and starting from the average level of relative permutation distances, the more $\alpha$ increases, the more weight is given to the highest relative permutation distances, and thus the higher the aversion to information loss. Notably, these two notions of aversion circumvent the issue involved in the empirical assessment of disclosure risk ([17]), where a score based on different measures of disclosure risk is computed using an ad-hoc weighting scheme. Under such an approach, weights can drive the overall assessment that is made. But using power means, measures can be computed on a continuum of weights which all carry an interpretation in terms of disclosure risk and information loss, thus providing a better grasp on the reality of micro data exchange, where various views about disclosure risk and information loss often meet.

Now, it must be noted that while these measures have also been originally proposed for the ex-post evaluation of disclosure risk and information loss ([11]), i.e. after having performed reverse-mapping for any method applied on any data set, in what precedes they have been purposely presented as ex-ante measures, i.e. by deriving absolute and relative permutation from the cipher's keys. In fact, it is one of the proposals of this paper to use power means as a guidance to calibrate the cipher's keys, as power means can be used equally effectively ex-ante or ex-post. However, before developing this notion, we provide a novel theoretical characterization of power means which, we believe, offers a powerful justification for their use.

### 4.2 A theoretical characterization of power means

Power means satisfies a set of basic properties and are already well-known outside the field of data anonymization ([20]). Here, and in the context of this paper, denoting a distribution of permutation distances by $p=(p_1,…,p_n)$, being *relative or absolute*, $J(p,\alpha)$, the power mean of parameter $\alpha$ for the evaluation of p, satisfies the following:

- ***Neutrality in evaluation (NE)****: if q is a permutation of p, then $J(q,\alpha)= J(p,\alpha)$*

This condition ensures that all the information used to evaluate p is considered equally.

- ***Size independence (SI)****: if $q=(p,p,…,p)$ is a m-duplicate of p (with m≥2), then $J(q,\alpha)= J(p,\alpha)$*

This condition connects the comparability of $J(p,\alpha)$ across data sets of different sizes, by establishing the ground for comparison on a per record basis.

- ***Normalization (NO)****: if $p_i= p_j=a$ for $i,j=1,…,n$, then $J(p,\alpha)=a$*

Normalization ensures that if all the permutation values in p are equal, then $J(p,\alpha)$ is equal to this permutation value.

- ***First degree homogeneity (FD)****: if $q=\lambda p$ for a scalar $\lambda>0$ $J(q,\alpha)= \lambda\, J(p,\alpha)$*

If the levels of permutation are magnified by the same scalar, so is the power mean.

- ***Continuity (CO)****: $J(p,\alpha)$ is continuous*

A standard assumption, continuity makes sure that the power mean doesn't change abruptly for small variations in p.

- **Sub-domain coherency (SC)**: For p' and p of the same size and q and q' of the same size, if J(p',α) > J(p,α) and J(q',α) = J(q,α), then J((p',q'),α) > J((p,q),α)

Sub-domain coherency establishes that if the absolute or relative permutation distances from two sub-data sets change in a way that leads to an increase in the power mean in one and remains unaltered in the other, then the overall power mean must increase. Stated otherwise, if absolute permutation distances increase in one sub-set but remain unchanged in the rest of the data set, then protection against disclosure risk must increase on the overall data set. Along the same lines, if relative permutation distances increase in one sub-set but remain unchanged in the rest of the data set, then information loss must increase in the overall data set.

The fact that the class of power means satisfies (NE), (SI), (NO), (FD), (CO), (SC) is trivial. However, less trivial is the fact that this is the only class of measures to do so:

**Theorem:** *An aggregative structure for the evaluation of disclosure risk and information loss satisfies (NE), (SI), (NO), (FD), (CO) and (SC) if and only if it is a power mean.*

**Proof:** *For necessity, we left the proof to the reader. For sufficiency, we start by assuming a function J(.) that satisfies (NE), (SI), (NO), (FD), (CO) and (SC). In what follows, permutation distances can be defined in relative or absolute terms indifferently.*

*Consider the universe of all possible data sets of at least 3 records, i.e. n≥3, and pick in this universe four of them which, after anonymization, generate four distributions of permutation distances: p and q of size m<n, and p' and q' of size m'=n-m. Then, assume that J(p,p') ≥ J(q,p'). (SC) precludes having J(p) < J(q), which thus implies J(p) ≥ J(q). If this inequality holds strictly, then by (SC) we have J(p,q') ≥ J(q,q'). But if inequality is not strict, then by (SC) J(p,q') < J(q,q') doesn't hold because J(p,q,q') < J(q,q',p) would contradict (NE). As a result, we have J(p,p') ≥ J(q, p') ⇒ J(p,q') ≥ J(q, q'). That means, bearing in mind that J(.) is assumed to verify (CO), that J(.) is strictly separable in every data set partition, which implies, following ([21]), that J(p) can be expressed as:*

$$J(p) = Z_n \left( \sum_{i=1}^{n} \Omega_n(p_i) \right)$$

*for every p of size n and with $\Omega_n(.)$ continuous and $Z_n(.)$ continuous and strictly increasing.*

*So far, what have been demonstrated is that (SC), (NE) and (CO) leads inevitably to a separable function. Now, what follows works along the same line as [22], which uses separabality to characterize power means.*

*By (NO) we have $a = Z_n(\sum_{i=1}^{n} \Omega_n(a))$ for a>0, which leads to $Z_n^{-1}(a) = n\Omega_n(a)$. Assuming $H_n = Z_n^{-1}(a)$ with $H_n(.)$ continuous and strictly increasing, J(p) can be rewritten as:*

$$J(p) = H_n^{-1}\left(\frac{1}{n}\sum_{i=1}^{n} H_n(p_i)\right) \text{ for every p of size n≥3}$$

*From this last equation assume $H = H_4$ and m=4n. We can write:*

$$H(J(p)) = H\left[H_m^{-1}\left(\frac{1}{m}\sum_{i=1}^{m} H_m(H^{-1}(H(p_i)))\right)\right]$$

$= \Theta_m^{-1}\left(\frac{1}{m}\sum_{i=1}^{m} \Theta_m(H(p_i))\right)$ *with $\Theta_m(.) = H_m(H^{-1}(.))$ strictly increasing and continuous*

*Once again, we have $\Theta_m(a) = a$ and in particular $\Theta_4(a)$. From here set p with n=2, p' its 2-duplicate and p'' its m-duplicate. (SI) implies (with in what follows $w_i = H(p_i)$):*

$$H(J(p'')) = \Theta_m^{-1}\left(\frac{1}{m}\sum_{i=1}^{m} \Theta_m(H(p_i''))\right)$$
$$= \Theta_m^{-1}(0.5 * \Theta_m(w_1) + 0.5 * \Theta_m(w_2)) = H(J(p'))$$
$$= \Theta_4^{-1}(0.5 * \Theta_4(w_1) + 0.5 * \Theta_4(w_2)) = 0.5 * (w_1 + w_2)$$

*Thus, $\Theta_m(.)$ must satisfy:*

$$0.5 * \Theta_m(w_1) + 0.5 * \Theta_m(w_2) = \Theta_m(0.5 * (w_1 + w_2))$$

*This last equation is a Jensen's functional equation having the following solution ([23]):*
$\Theta_m(b) = a_m * b + c_m$ *for some scalars* $a_m$ *and* $c_m$.
*This solution implies for m=4n:*

$$H(J(p)) = \frac{1}{m}\sum_{i=1}^{m} H(p_i)$$

*Now, for a given data set with n≥1 and its four-duplicate, with p and p' the respective distribution of permutation distances, it holds by (SI) that*

$$H(J(p)) = H(J(p')) = \frac{1}{n}\sum_{i=1}^{n} H(p_i) = \frac{1}{4n}\sum_{i=1}^{m} H(p'_i)$$

*In turn, this implies that:*

$$J(p) = H^{-1}\left[\frac{1}{n}\sum_{i=1}^{n} H(p_i)\right]$$

*Now, consider a data set with two observations and a scalar $\vartheta > 0$. By (FD) and the equation above it holds that (with in what follows $w_i = H(p_i)$, meaning that $H^{-1}(w_i) = p_i$):*

$$H[\vartheta H^{-1}(0.5 * H(p_1) + 0.5 * H(p_2))] = 0.5 * H(\vartheta p_1) + 0.5 * H(\vartheta p_2)$$
$$\Rightarrow H[\vartheta H^{-1}(0.5 * w_1 + 0.5 * w_2)] = 0.5 * H(\vartheta H^{-1}(w_1)) + 0.5 * H(H^{-1}(w_2))$$
$$\Rightarrow H^\vartheta[H^{-1}(0.5 * w_1 + 0.5 * w_2)] = 0.5 * H^\vartheta(\vartheta H^{-1}(w_1)) + 0.5 * H^\vartheta(H^{-1}(w_2))$$

with $H^\vartheta(a) = H(\vartheta a)$ for a>0
*Now, assuming* $L^\vartheta(a) = H^\vartheta(H^{-1}(a))$ *we have:*

$$L^\vartheta(0.5 * w_1 + 0.5 * w_2) = 0.5 * L^\vartheta(w_1) + 0.5 * L^\vartheta(w_2)$$

*Following ([23]) the solution to this Jensen's functional equation is:*
$L^\vartheta(b) = x^\vartheta * b + y^\vartheta$ *for some scalars* $x^\vartheta$ *and* $y^\vartheta$.
*Now, using H(b)=a it holds that:*

$$H(\vartheta b) = x(\vartheta)H(b) + y(\vartheta)$$

*Following ([24]) the solution to this functional equation is:*

$$H(b) = \begin{cases} g * b^\alpha + h \text{ for } \alpha = 0 \\ g * \ln b + h \text{ for } \alpha \neq 0 \end{cases}$$

*But given that* $J(p) = H^{-1}\left[\frac{1}{n}\sum_{i=1}^{n} H(p_i)\right]$ *we thus have:*

$$J(p,\alpha) = \begin{cases} \left(\frac{1}{n}\sum_{i=1}^{n} p_i^\alpha\right)^{\frac{1}{\alpha}} \text{ for } \alpha \neq 0 \\ \prod_{i=1}^{n} p_i^{\frac{1}{n}} \text{ for } \alpha = 0 \end{cases}$$

*Thus, $J(p,\alpha)$ is a power mean, which completes the proof.*

This result establishes power means as the only aggregative structure which, alongside a set of standard properties, satisfies sub-domain coherency. It is a result valid beyond the context of data anonymization, in fact for any vector of any quantities to be evaluated. It must also be emphasized that power means have been previously theoretically characterized in the literature ([22]), but by postulating at the onset the condition of separability. The result in this paper extends this previous work by demonstrating that separability appears to be in fact based on three conditions: neutrality in evaluation, continuity and sub-domain coherency. It is this last one that is of particular and practical importance for data anonymization, as it turns out that only power means can coherently cope with anonymization by block of records.

### 4.3 Ex-ante calibration of permutation and a new approach to data anonymization

As stated earlier, data anonymization is currently practiced using a variety of methods, often very heterogeneous in nature and with some of them now very well-established in the literature. However, regardless of the many choices available, at a general level they are all used the same way (Figure 1). A method is selected with the anonymization practitioner having in mind either a utility-first or a privacy-first approach, and is applied to a data set. The outcome of this is then evaluated using specific measures of disclosure risk and information loss. But as mentioned earlier, because the methods' parameters in themselves are a poor guide to inform about the final levels of privacy and information obtained, as for a given parametrization different outcomes are possible according to the distributional features of the data, a necessary and specific ex-post checking step leads generally to some re-runs before reaching an anonymized version of the data viewed as acceptable. Additionally, because the ex-post checking is specific, the comparison of performances across different methods is an arduous task ([9]).

**Figure 1. Current approach to data anonymization**

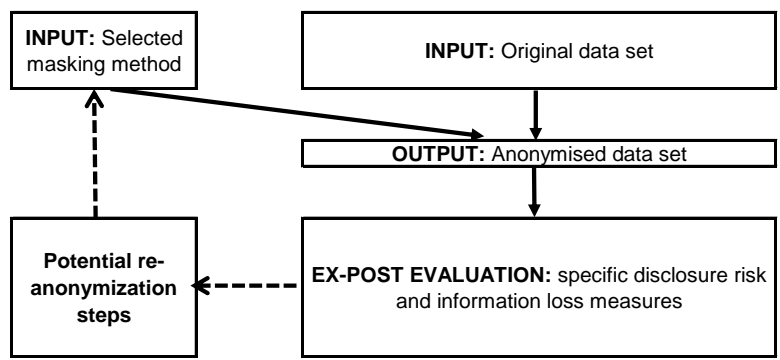

We saw already that the use of power means on absolute and relative permutation distances provides a ground for universal ex-post checking, based on the retrieval of the permutations pattern that a method has generated. But at the conceptual level, the fact of using a method that unavoidably leads to a permutations pattern (plus eventually but unnecessarily a small noise addition), or applying this permutation pattern directly by using the cipher previously developed, is equivalent. These two ways will lead strictly to the same outcome in terms of risk and information. However, the latter appears to be more efficient, as once the permutations pattern has been set, it will be automatically translated into the final, anonymized data set. In fact, this will avoid the empirical ex-post checking stage and some eventual iteration to attain the desired levels of disclosure risk and information loss. This leads to a new approach for the practice of individual data anonymization (Figure 2).

**Figure 2. New approach to data anonymization**

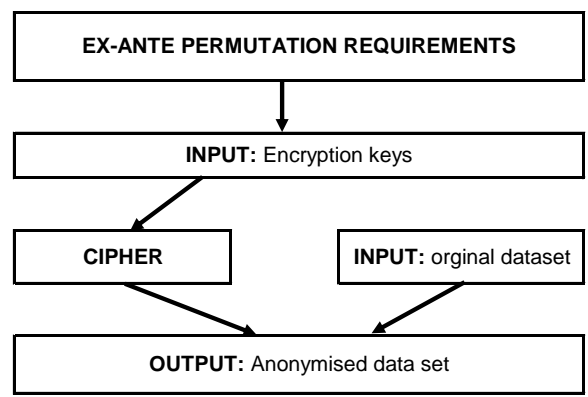

Of course, for this new approach to be practical, it requires thinking about anonymization only in terms of permutation. The permutation paradigm already pointed out that any anonymization method is equivalent to applying permutations. This is in a way a new language for data anonymization. With classical methods it is primarily their parameters (for example the variance for noise addition), and their varying strengths the language, which allow translating some targeted levels of disclosure risk or information loss into practice, albeit due to the varying nature of the data this translation is rarely perfect at the end. Now to set permutation as a language to perform anonymization ex-ante, it is needed to expand its vocabulary so as to provide guidance on how to build the cipher's keys.

As we saw in Section 2, generally a data exchange requires two groups of agents: a data provider and the data users. The former wants to disseminate some individual data for some users that are in need of them. But prior to the exchange the provider, equipped with some raw, non-anonymized data, needs to secure them so that no individuals could be reasonably identified, while at the same time providing an acceptable level of information. To achieve this he will undertake data anonymization himself. Now, let's introduce a new third agent, the permutation provider, whose task is to build some suitable permutation keys. Clearly, this new agent will never need to see the data. He can just work in isolation on the keys, having as information the number of attributes and individuals in the data, signaled by the releaser. However, what the releaser has to do is formulate some desiderata on how he wants the data to be anonymized. This can be expressed through a permutation menu.

First, and for disclosure risk, the data releaser must advise the amount of permutation for each attribute. For example, for a given attribute, he can advise that he wants all records at least permuted one time, while at the same time a certain average of permutations must be achieved. For other attributes, these constraints can be modified, for example not all individuals must be permuted, or the average amount of permutation can be lower or reinforced, for example every individual must be permuted at least two times and the average amount of absolute permutation must be high. Second, and for information loss, the releaser must notify which couple of attributes are critical in terms of information and must be preserved to a large extent, with a small average of relative permutation distance. The other less valuable couples in terms of information can then be relatively permuted higher on average or within a certain portion of the distribution of relative permutation distances. Obviously, all the requirements in a permutation menu must be formulated simultaneously, as the keys taken in isolation make up for disclosure risk, while it is their relative properties taken by pair that make up for information loss. The data releaser must then formulate all his demands simultaneously to the permutation provider and must pay attention to the coherency of his requests, bearing in mind for example that two attributes cannot be protected with very dissimilar keys if at the same time their joint distribution has to be reasonably preserved. Keeping up with such coherency simply means coping with the unavoidable protection/information trade-off in data anonymization.

Now, power means constitute a way to create a permutation menu. For different scenarios of risk and information aversion, different levels of power means can be required ex-ante, from which the permutation provider will reconstitute the permutation keys. Of course, technically speaking it is clear that there may be no unique way to create permutation matrices from various values of power means. This won't affect the overall level of protection and information for the anonymized data set, while of course it could change the property of verifiability by the subjects ([2]): for a given set of power means values and the associated levels of protection of information, different keys could lead to a given individual being permuted differently. This is, however, a minor issue. There may be also no permutation keys that can be derived from a set of power means, but this problem can be avoided first by ensuring the coherency of the permutation menu proposed.

While power means is one way of creating a permutation menu for then generating keys, it must be recognized that there may be other ways. However, we just saw that power means are the only measures that are sub-domain coherent, which is a powerful justification for using them. Notably, and as far as big data are concerned, it can offer some obvious practical benefits. For instance, anonymization can be performed by blocks to ease the computational workload: when the data are split in m blocks, with some given levels of protection and information on m-1 blocks, then the anonymization of the $m^{th}$ block will lead to an increase in protection of the *overall* data set. Such coherency won't be ensured by other measures.

## 4.4 Examples of permutation menus

We now provide some empirical examples of permutation menus. To make the bridge with existing methods, we will first show the menus uncovered by some of them. We will then turn to menus conceived independently of any method, i.e. based on power means guidance only.

Our first examples retrieve the permutation menus of one rank-based method, i.e. rank swapping, and two noise-based methods, i.e. additive and multiplicative noise injection. The experimental data set used is two attributes of the Census data set observed over 1080 records. This data set has been used several times in the literature to evaluate the properties of anonymization techniques in terms of disclosure risk and information loss ([25]). Additive noises are injected with a standard deviation equal to 50% of the standard deviations of the two attributes, multiplicative noises are drawn from a uniform distribution within the range (0.75,1.25) and rank swapping is set with a swapping distance of 30%. Masked data are then reverse-mapped to compute the levels of absolute and relative permutations. From these levels, we then compute the universal measures of disclosure risk and information outlined above for a quasi-continuum of aversion parameters, i.e. by increments of 0.01. As suggested in ([11]), the results are finally displayed directly under the form of curves with the aversion parameters on the x-axis and the measures of disclosure risk for each attributes and the measures of information loss for the couple of attributes on the y-axis. These graphical representations display the permutation menus for all ranges of aversion to disclosure risk (Figures 3 and 4) and information loss (Figure 5).

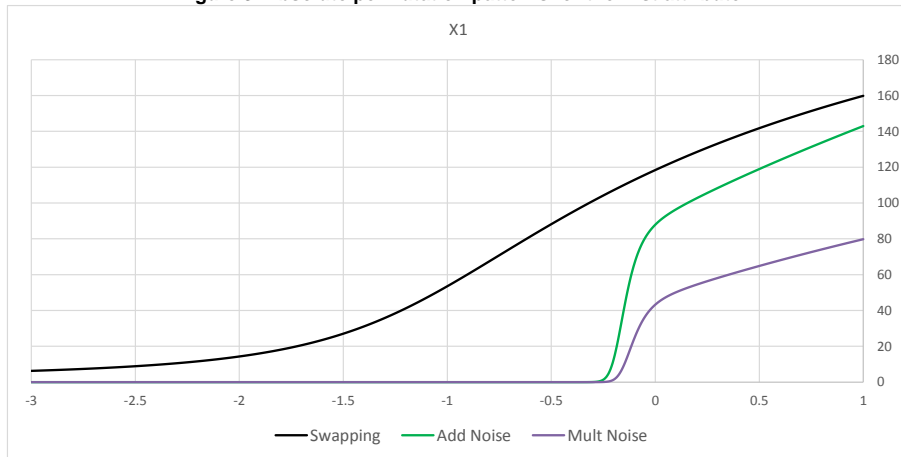

Figure 3. Absolute permutation patterns for the first attribute

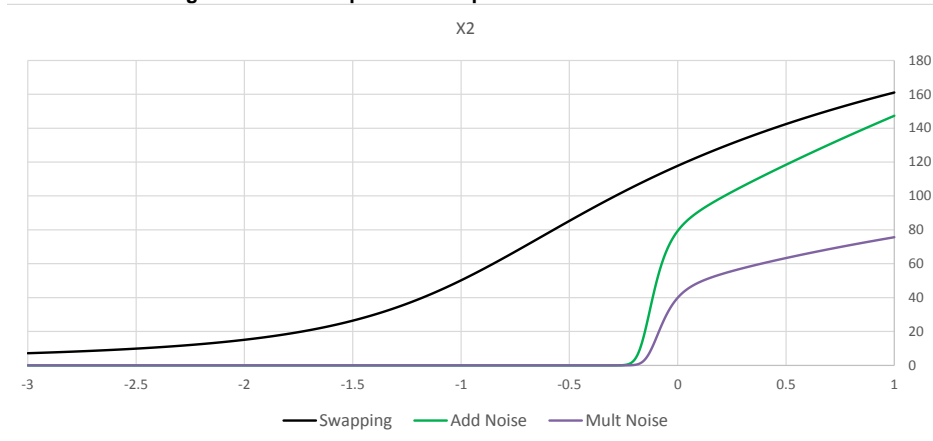

Figure 4. Absolute permutation patterns for the second attribute

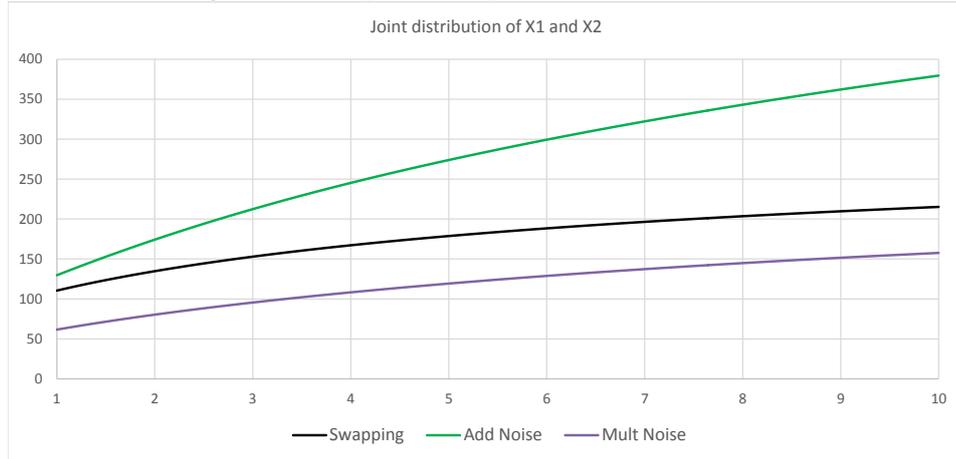

Figure 5. Relative permutation patterns for the two attributes

For the two attributes, the three permutation menus proposed differ across the methods considered. Data swapping is equivalent to formulating the following requirements[5]:

- All records must be permuted at least one time (as the associated curves in Figure 1 and 2 never reach zero).
- A large portion of the records must be permuted at least 20 times (as the associated curves in Figure 1 and 2 are above 20 from an aversion around -1.7 and above).
- On average each record must be permuted 160 times (which is the value reached by the curves when $\alpha=1$, i.e. the average levels of absolute permutation distances).

However, noise-based methods deliver the following menus:

- Not all records have to permuted and in fact a large portion of them can remain untouched for the two attributes (as the associated curves in Figure 1 and 2 quickly hit zero).
- On average each record must be permuted 80 or 140 times.

Regarding information loss (Figure 5), the permutation menus can be described as follow:

- On average the relative permutation distances can be relatively high (around 120 for noise addition and swapping) or low (60 for multiplicative noise).
- Potentially, a large portion of records can be relatively permuted high (for noise addition the maximum relative permutation distance is established at almost 400).
- As a result, additive noise proposes some permutation requirements for which the constraint of preserving information between the two attributes (and thus the similarity between the permutation keys of each attributes) is looser than the requirements expressed by the swapping and multiplicative noise methods.

Now, instead of starting from already existing methods, one can start from some permutation menus and the associated curves and then generate the permutation keys that will fulfill these requirements. Let's assume the following:

---

[5] Here the menus expressed are using permutation distances as the unit. By normalizing with the maximum permutation distances possible in the data set (and thus re-scaling Figure 3, 4 and 5), they can also be expressed in percentages, with the power means values ranging between 0 and 1 ([11]).

- For the first attribute, we require that all records must be permuted at least one time and that the average level of absolute permutation must be high (menu 1). Alternatively, we require a low level of average absolute permutation in conjunction with a large chunk of records not being permuted (menu 2).
- For the second attribute, we require quite similar menus with a large chunk of records not permuted at all, albeit we also set menu 1 to have an average level of absolute permutation almost twice as high than menu 2.
- As a result, we aim at two different scenarios for information loss. With menu 1 the keys for the two attributes are relatively dissimilar in their profiles, not least because the first key must permute all records while the other not. However, with menu 2 the keys are relatively similar. Consequently, we relax on purpose the constraint of information preservation for menu 1 while menu 2 must preserve it to a large extent.

Figure 6, 7 and 8 display the resulting permutation requirements when one starts from power means desiderata, creates the associated vectors of absolute and relative rank displacements and then generates the underlying permutation matrices. Notably, one can see that in the second menu relative permutation distances are small for whatever scenario of aversion to information loss, while this is the contrary for the first menu (Figure 8). This result is ensured by the similar absolute permutation profiles for the two attributes requested in menu 2 (Figures 6 and 7). Now, when thinking about data anonymization only in terms of permutation as a universal approach, as we just did, the data can then be anonymized using the created keys and the cipher of *Proposition 3*. The ex-post properties in terms of disclosure risk and information loss will be strictly the same as the ones determined ex-ante.

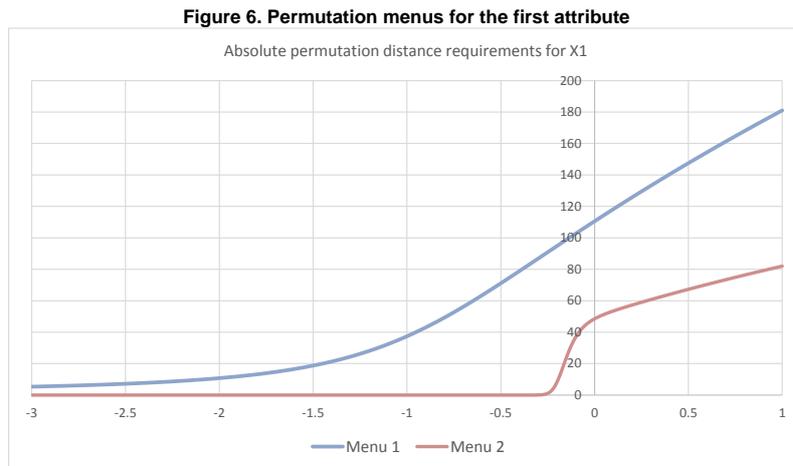

**Figure 6. Permutation menus for the first attribute**

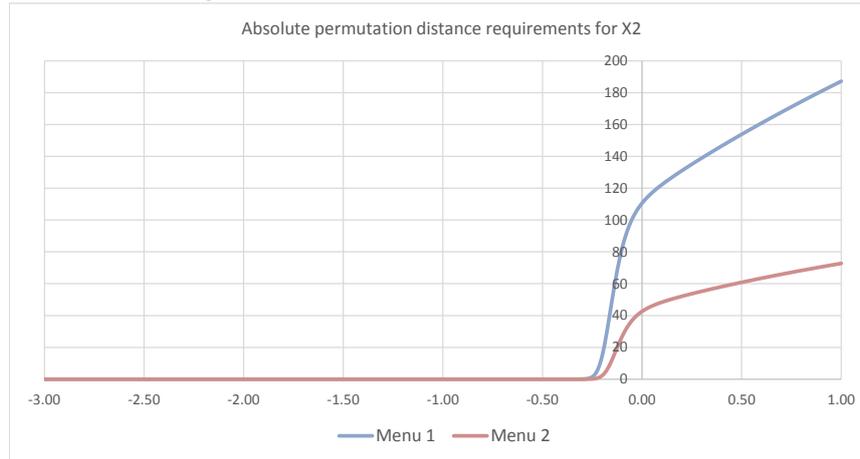

**Figure 7. Permutation menus for the second attribute**

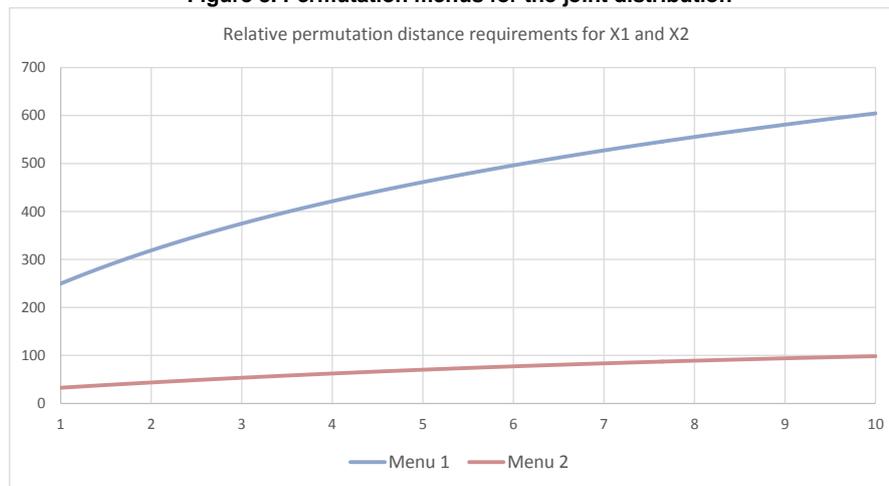

**Figure 8. Permutation menus for the joint distribution**

## 5. Conclusions and future research

In ([2]) a new anonymization paradigm, which views any methods as functionally equivalent to permutation, was presented for the evaluation of any anonymization methods applied on any data set, while in ([11]) a set of universal metrics for disclosure risk and information loss were developed based on this paradigm. These developments were not seeking for a new anonymization framework *per se*, but instead tried to establish an analytical environment for the comparison of currently existing methods in a sound and universal way. In this paper, we have challenged this limitation of scope by arguing that these results can be as effective pre-anonymization as they are post-anonymization. Borrowing from cryptography, we have developed for the first time a general cipher for data anonymization. This cipher is able to replicate the outcome of any method, and some of its properties outline general lessons for data anonymization. In particular, at a general level of functioning, anonymization can always be performed independently of the data to be anonymized. As a result, beyond being a universal mimicker, the cipher is a tool in itself that can be used through the exploration of permutation structures. We then provided some guidance about how to explore these structures, notably by proposing to calibrate permutation keys using power means, for which we also suggested a new theoretical justification. The tools proposed in this paper allow for a more efficient, ex-ante approach to data anonymization.

We leave as future work the deeper exploration of these proposals. First of all, the generation of permutation through the reverse engineering of power means should be enhanced and established further. Second, as it appears that data anonymization relies on the single principle of permutation, which could be phrased as a general principle as "to be protected, become someone else", an intuitive privacy guarantee and thus a new privacy model should be developed around such a principle. Third, the exploration of the composition of an approach by permutation is warranted, i.e. when merging two data sets with certain permutation patterns, identifying the result of the merge with its subsequent privacy and information guarantees.